\documentclass[aps,prl,twocolumn,showpacs,preprintnumbers,amsmath,amssymb,groupaddress]{revtex4}

\usepackage{graphicx}
\usepackage{dcolumn}
\usepackage{bm}

\begin{document}
\preprint{M. W. Kim \textit{et al}.}

\title{Enhanced Optical Gap in Bi-layered Manganites La$_{2-2x}$Sr$_{1+2x}$Mn$_{2}$O$_{7}$ near $x=0.4$}
\author{Myung Whun Kim$^{1}$}
\author{H. J. Lee$^{1}$}
\altaffiliation[current address: ]{Department of Physics,
University of California at Berkeley, Berkeley, CA 94720 USA. }
\author{B. J. Yang$^{2}$}
\author{Kee Hoon Kim$^{2}$}
\author{Y. Moritomo$^{3}$}
\author{Jaejun Yu$^{2}$}
\author{T. W. Noh$^{1}$}
\email[corresponding author: ]{twnoh@snu.ac.kr}
\affiliation{$^{1}$ReCOE \& School of Physics, Seoul National
University, Seoul 151-747, Korea\\
$^{2}$CSCMR \& School of Physics, Seoul National University, Seoul
151-747, Korea\\
$^{3}$Department of Physics, University of Tsukuba, Tsukuba
305-8573, Japan}
\begin{abstract}
We have systematically investigated the optical conductivity
spectra of La$_{2-2x}$Sr$_{1+2x}$Mn$_{2}$O$_{7}$ ($0.3 \leqslant x
\leqslant 0.5$). We find that just above the magnetic ordering
temperatures, the optical gap shows an enhancement up to $\sim
0.3$ eV near $x=0.4$. Based on a $x$-dependent comparison of the
nesting vector of the hypothetical Fermi surface and the
superlattice wave-vector, we suggest that the peculiar
$x$-dependence of the optical gap can be understood in terms of
charge and lattice correlation enhanced by the charge density wave
instability in nested Fermi surface.
\end{abstract}
\maketitle

The striped charge and spin correlation mark one of the generic
features of doped Mott insulators with strong electron
correlation. In particular, there exist numerous examples of the
enhanced charge/spin correlation at commensurate hole doping,
often signaled by an increased charge gap and anomalies in the
spin/lattice degree of freedom. The famous $y=1/8$ anomaly and
stripe correlation in La$_{1.6-y}$Nd$_{0.4}$Sr$_y$CuO$_4$
\cite{Tranquada}, the static charge-spin stripe and an enhanced
optical gap near $y=1/3$ of La$_{2-y}$Sr$_{y}$NiO$_{4}$
\cite{Katsufuji,Yoshizawa}, and the $CE$-type charge/orbital order
with a large optical gap near $y\sim 1/2$ of
La$_{1-y}$Ca$_y$MnO$_3$ \cite{KH}, all constitute well-known cases
of the unusual stability of charge/spin correlation at the
commensurate dopant levels. Although the anomalies at the specific
commensurate dopant concentrations are fascinating in their own
right, they are still not well understood. Furthermore, dynamic
and/or short-ranged striped correlations remain as quite
challenging issues in strongly correlated electron systems. They
often exhibit themselves as an intriguing coexistence and coupling
among charge/spin/lattice correlation at nanoscopic length scales.

La$_{2-2x}$Sr$_{1+2x}$Mn$_{2}$O$_{7}$ has exhibited an intriguing
coexistence of striped charge/lattice correlations near or above
the long range magnetic ordering temperature
\cite{Vasiliu-Doloc,Campbell}. From the early stages of the
development of this field, the origin of the strong localization
tendency of the doped holes has been a subject of continual
interest and debates \cite{Littlewood}. While the system is
supposed to be a ferromagnetic (FM) metal at low temperatures,
high resistivity \cite{Moritomo-Nature} and non-Drude type optical
conductivity spectra \cite{HJ} indicate that it should be close to
a localized state. A recent angle-resolved photoemission (ARPES)
experiment reported that a minimal Fermi surface (FS) of $x=0.4$
is formed below $T_{\rm C}$ along the nodal direction while most
of the 2D FS is gapped \cite{Mannella}. [Note that this
anisotropic FS is quite similar to that of high $T_{\rm C}$
superconductors.] While strong electron-phonon interaction has
been hypothesized to be a source of the localization, it is not
clear how such a bosonic interaction microscopically influences a
polaronic metal system with an anisotropic band structure to tune
localized states.

Herein, we report experimental findings on the enhanced optical
gap in La$_{2-2x}$Sr$_{1+2x}$Mn$_{2}$O$_{7}$ (LSMO) near $x=0.4$
through systematic doping dependent studies of optical
conductivity spectra. The enhancement of the optical gap occurs at
unusual specific hole doping, \textit{i.e.} at $x=0.4$, which is
clearly distinguished in the previous well-known cases. Based on
the comparison between the doping-dependent nesting vector of the
FS and $k$-vector of the superlattice coming from the
charge/lattice correlation, we suggest that charge density wave
(CDW) instability in the nested FS plays a key role in causing the
enhancement of the optical gap near $x=0.4$.

Single crystals of La$_{2-2x}$Sr$_{1+2x}$Mn$_{2}$O$_{7}$ ($0.3
\leq x \leq 0.5$) were grown by the floating-zone method. The
samples were characterized by resistivity and magnetization
measurements \cite{Moritomo-Nature}. For optical reflectivity
measurements, cleaved $ab$-planes were prepared. Temperature ($T$)
dependent reflectivity spectra $R(\omega)$ ($\omega$ is the photon
energy) were measured at a temperature range of 15 - 300 K and
over a wide $\omega$ range of 5 meV - 30 eV. The NIM-Beam line at
Pohang Accelerator Lab was used for the high energy (5 - 30 eV)
measurement. We used the Kramers-Kronig transformation to obtain
the optical conductivity spectra ($\sigma(\omega)$) from
$R(\omega)$. Details of the optical experiments were described in
a previous paper \cite{HJ}.

\begin{figure}[tbp]
\includegraphics[width=3.2in]{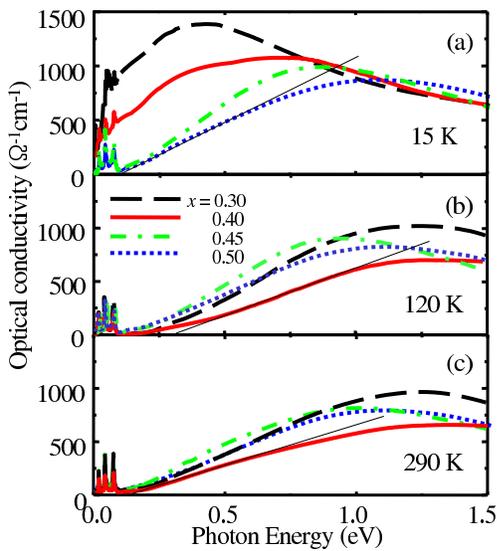}
\caption{(color online). $\sigma(\omega)$ of
La$_{2-2x}$Sr$_{1+2x}$Mn$_{2}$O$_{7}$ ($0.3 \leq x \leq 0.5$) (a)
at 15 K, (b) at 120 K, and (c) at 290 K. The thin solid lines
represent the extrapolation for 2$\Delta$ estimate.}\label{Fig:1}
\end{figure}

LSMO shows different magnetic ground states at different $x$
\cite{phase}. For $x \sim 0.3$, the spins in the MnO$_2$ bilayers
show FM alignment along the $c$-axis. For $0.33 \leq x \leq 0.4$,
the FM spins align parallel to the $ab$-plane. For $0.4 \leq x
\leq 0.5$, the alignment between MnO$_{2}$ layers becomes canted.
Finally, $A$-type antiferromagnetic (AFM) order is stabilized for
$x\sim 0.5$. We note that in all the compounds with $0.3 \leq x
\leq 0.5$, the spins within a single MnO$_{2}$ layer are always
ferromagnetically aligned below $T_{\rm C}$ or $T_{\rm N}$,
regardless of their long-range ordering pattern.

In Fig. 1 (a), $\sigma(\omega)$ of both $x = 0.45$ and 0.50 at
$T=15$ K are clearly suppressed below 1.0 eV, forming a finite
optical gap. It is known that charge ordering (CO) instability
occurs in $x=0.50$ below 210 K, although it becomes unstable and
short-ranged below 100 K with the development of the $A$-type AFM
state \cite{Argyriou}. Therefore, the spectral features are
consistent with the localization of carriers in short-ranged CO
states for both compounds. We can define an optical gap
($2\Delta$) as an onset energy of the steeply rising part of
$\sigma(\omega)$ determined from a crossing point between the x
abscissa and a linear extrapolation line drawn at the inflection
point of $\sigma(\omega)$ (thin solid lines in Fig. 1), following
the common practice used to evaluate $2\Delta$ of various CO
materials \cite{KH}.

On the other hand, $\sigma(\omega)$ of $x = 0.30$ and $0.40$ with
the FM metallic ground state show broad maxima around 0.5 and 0.7
eV, respectively. The broad maximum, unexpected within the
conventional Drude model, is the absorption due to the incoherent
hopping motion of carriers from Mn$^{3+}$ to Mn$^{4+}$ sites
\cite{MW,HJ}. It is interesting that the existence of the
$\sigma(\omega)$ maximum is accompanied by a decreasing
$\sigma(\omega)$ with $\omega\rightarrow 0$, resembling a gap
feature observed in $x=0.50$. We can attribute the decreasing
behavior of $\sigma(\omega)$ to the \textit{signature of the
pseudo-gap} even in the metallic states. Furthermore, the overall
spectral shape in Fig. 1(a) suggests that the nature of the
pseudo-gap might be directly related to the optical gap found in
the $\sigma(\omega)$ of $x=0.50$ and 0.45 at $T=15$ K.

In Fig. 1(b), $\sigma(\omega)$ at 120 K reveal yet another
intriguing finding. $2\Delta$ ($\sim 0.3$ eV) is clearly enhanced
at the specific compound $x=0.40$. It is even larger than that of
$x=0.50$ at which the CO state is stabilized. Such an enhanced
$2\Delta$ for $x=0.40$ seems to exist even up to 290 K, as shown
in Fig. 1 (c). The $x$-dependent spectral response above $T_{\rm
C}$ is counterintuitive because the $2\Delta$ of $x=0.40$ with a
metallic ground state is clearly larger than that of $x=0.50$ with
the CO insulating state.

To understand the unexpected $x$-dependence of $2\Delta$ at 120 K,
we have systematically investigated $T$-dependence of $2\Delta$
for all the samples. As shown in Fig. 2(a), $T-$dependence of
$2\Delta$ confirms its anomalous enhancement near $x=0.40$. First,
$2\Delta$ of both $x = 0.45$ and 0.50 compounds is only about 0.1
eV, and show a weak $T$-dependence. Furthermore, $2\Delta$ of $x =
0.50$ even slightly decreases below 100 K in contrast to one of
its 3D analogues \cite{note}. On the other hand, in the compounds
with $x = 0.30, 0.40,$ and $0.42$, in which the 3D FM metallic
ground states are stabilized, the $2\Delta$ above $T_{\rm C}$ is
larger than that of $x=0.45$ or 0.50. As displayed in Fig. 2(b),
$2\Delta$ values at just above $T_{\rm C}$ or $T_{\rm N}$
systematically increase from 0.1 (for $x=0.50$) and 0.3 (for
$x=0.40$), and then decrease again to 0.2 eV (for $x=0.30$).

\begin{figure}[tbp]
\includegraphics[width=3.2in]{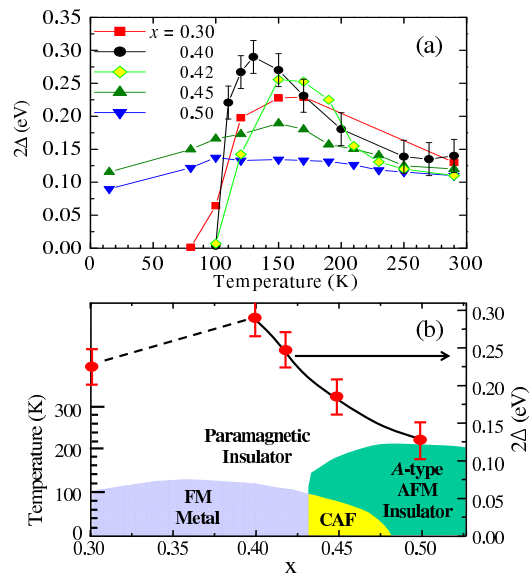}
\caption{(color online). (a) Temperature dependent $2\Delta$
obtained from the $\sigma(\omega)$. The error bars of other
compounds are almost the same size as that of $x=0.40$. (b)
$x$-dependence of $2\Delta$ obtained at just above $T_{\rm C}$ or
$T_{\rm N}$ is overlaid on the phase diagram. The lines are guides
to the eyes.}\label{Fig:2}
\end{figure}

The $2\Delta$ value of $x=0.50$ shows a decreasing behavior as $T$
drops below around 100 K, as shown in Fig. 2(a). It is consistent
with a previous neutron scattering experiment, which suggested the
melting of CO correlation below 100 K with the stabilization of
the 2D FM state \cite{Argyriou}. Therefore, it is likely that CO
correlation is rather weak at overall $T$ in $x=0.45$ and 0.50,
consistent with the weak or decreasing $2\Delta$ behaviors shown
in Fig. 2(a). On the other hand, the enhanced $2\Delta$ near
$x=0.40$ above $T_{\rm C}$ gives rise to a surprising observation
that the localization tendency could be at its largest for
$x=0.40$ just above its $T_{\rm C}$ while the system has a FM
metallic ground state consistent with the collapse of $2\Delta$
below $T_{\rm C}$.

The enhancement $2\Delta$ of above $T_{\rm C}$ in $x=0.40$ is
reminiscent of the similar behavior of $2\Delta$ observed in other
3D manganite systems, $i.e.$ La$_{1-y}$Ca$_{y}$MnO$_{3}$ near
$y=0.50$: $2\Delta$ at $T=10$ K systematically increases from 0.1
to 0.5 eV as $y$ decreases from 0.80 to 0.50. The enhancement of
the CO gap at $y=0.50$ is due to the unusual stability of the
$CE$-type CO configuration at the commensurate doping. The finite
$2\Delta$ up to far above $T_{\rm CO}$ has been attributed to the
short range, and the $CE$-type CO correlation resulting from the
suppression of $T_{\rm CO}$ near a thermodynamic bi-critical point
at $y=0.50$ \cite{KH}.

However, the finite and enhanced $2\Delta$ behaviors above $T_{\rm
C}$ in LSMO near $x=0.40$ cannot be simply understood as a result
of the proximity to the bi-critical point. As inferred from the
phase diagram in Fig. 2(b), the possible bi-critical point, if
any, should be rather close to $x=0.45$; however, $2\Delta$ of
$x=0.45$ shows much weaker $T$-dependence than that of $x=0.40$.
In addition, CO stability seems to be weak in the present system
at overall $x$. The CO at the commensurate doping of $x=0.50$
appears in a narrow $T$-range (100 K $<T<$ 210 K), and even
diminishes below 100 K with the AFM ordering \cite{Argyriou}.
Therefore, we argue that the origin of the enhanced $2\Delta$ near
$x=0.40$ should be attributed to a new mechanism beyond the
$CE$-type CO correlation.

A recent ARPES experiment has provided compelling evidence that FS
of $x=0.40$ is formed below $T_{\rm C}$ only along nodal direction
while most of the 2D FS is gapped \cite{Mannella}. The resultant
FS is highly nested with the nesting vector $q_{F}\sim 0.3$, which
is susceptible to CDW formation, as suggested by Chuang \textit{et
al} \cite{Dessau}. The possibility of the CDW in $x=0.40$ has been
corroborated by x-ray scattering experiments
\cite{Vasiliu-Doloc,Campbell}; the stripe-like superlattice
modulation vector $q_{L}\simeq 0.3$ matches with the $q_{F}$
\cite{Dessau}.

The $T$-dependent $2\Delta$ of the $x = 0.40$ in Fig. 2(a) is also
consistent with CDW formation. Under CDW instability, the square
of $2\Delta$ develops in proportional to the superlattice peak
intensity, $I$, that results from the periodic lattice modulation
\cite{Gruner}. Strikingly, we found a close overlap between the
square-root of $I$ for the $(0.3,0,1)$ peak, observed by x-ray
scattering experiments \cite{Vasiliu-Doloc,Campbell} and the
optical gap: \textit{i.e.}, $\sqrt I \varpropto 2\Delta$, as
demonstrated in Fig. 3(a). Therefore, our optical study strongly
suggests the formation of CDW instability in the $x = 0.40$.

Why, then, is CDW instability enhanced near $x=0.40$? With the
lack of $x$-dependent experimental FS data, we have theoretically
investigated the FS topology of a MnO$_{2}$ layer. We used a
tight-binding method, under the assumption that the local Mn spins
are ferromagnetically aligned in a MnO$_{2}$ plane and $e_{g}$
orbitals are doubly degenerate. Figs. 4(a) and 4(b) show the
calculated FS, and Fig. 4(d) shows $x$-dependence of $q_{F}$ from
$x=0.01$ to 0.5. In Fig. 4(c) we present the FS of $x=0.40$, when
spins are not polarized. The calculations reveal a couple of
important findings: (1) The spin ordering is essential for the
formation of the nested FS (Compare Fig. 4(b) with 4(c)). (2) The
predicted $q_{F}$ decreases from 0.50 (for $x=0.01$) to 0.25 (for
$x=0.5$), with $q_{F}\approx0.33$ for $x=0.40$ being consistent
with ARPES results \cite{Mannella,Dessau}.

\begin{figure}[tbp]
\includegraphics[width=3.2in]{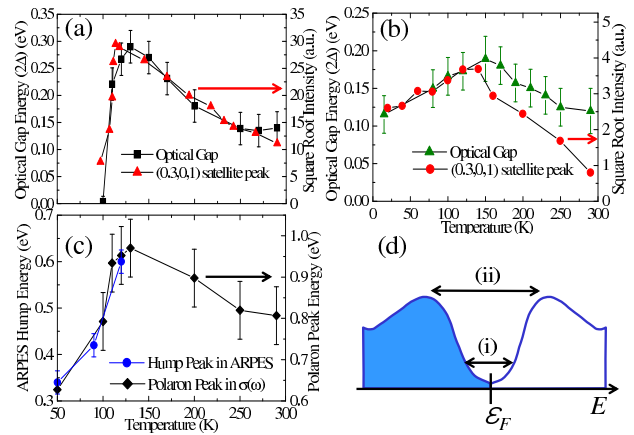}
\caption{(color online). Comparison between $2\Delta$ and $\sqrt
I$ of ($0.3,0,1$) superlattice peak (a) for $x=0.40$ and (b) for
$x=0.45$. $\sqrt I$ values were obtained from Ref.
\cite{Vasiliu-Doloc}. (c) Comparison between the hump peak energy
in the ARPES spectra (Ref. \cite{Mannella}) and the incoherent
absorption peak in $\sigma(\omega)$ as determined by Lorentz
oscillator fittings. (d) Schematic band diagram near the Fermi
energy ($\epsilon_{\rm F}$): (i) the pseudo-gap and (ii) the
incoherent absorption (polaron) peak
energy.}\label{Fig:3}\end{figure}

In contrast to the decreasing $q_{F}$ in Fig. 4(d), the
experimentally available points of $q_{L}$ slowly increase as $x$
increases in $0.3 \leq x \leq 0.5$, obtained from x-ray scattering
experiments \cite{Vasiliu-Doloc,Kubota}. It should be noted that
in Fig. 4(d) the $x$-dependent $q_{F}$ and $q_{L}$ curves cross
near $x=0.40$. When $x$ deviates from 0.40, the discrepancy
between $q_{F}$ and $q_{L}$ actually increases, which suggests a
scenario in which CDW instability could be mostly stabilized near
$x=0.40$ because of the close matching condition of $q_{F}$ and
$q_{L}$. Furthermore, in contrast to a simple 1D CDW case, $q_{F}$
and $q_{L}$ seem decoupled in this 2D system, suggesting that the
coincident matching of the two vectors near $x=0.40$ could have
driven the stabilization of the CDW correlation due to the
formation of the nested FS.

The CDW for other $x$, if formed, should be weaker than that of
$x= 0.40$. This seems to be indeed consistent with the decreasing
$2\Delta$ behavior in Fig. 2(b) as $x$ deviates from 0.40.
Furthermore, the comparison between $2\Delta$ ($T$) of $x=0.45$
and $\sqrt I$ of the superlattice peak of $x=0.44$ suggests that
they are a better match with each other near $T_{\rm C}$ and
below, as shown in Fig. 3(b). This finding supports the
calculations in Fig. 4 that CDW stability can be achieved near
$x=0.40$ under FM spin ordering.

\begin{figure}[tbp]
\includegraphics[width=3.2in]{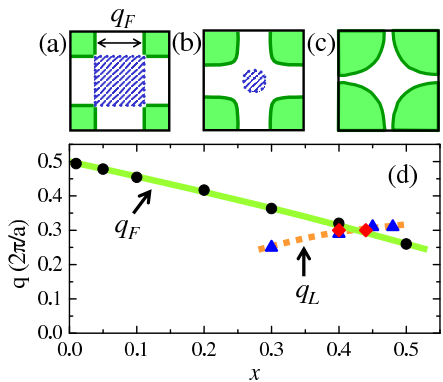}
\caption{(color online). The hyphothetical Fermi-surface of (a)
$x=0.01$, (b) 0.40 with FM spin ordering, and (c) $x=0.40$ without
the FM ordering (d) The solid circles represent the $x$ ependence
of $q_{F}$ with FM spin ordering, and the solid triangles and
solid diamonds represent the $q_{L}$, superlattice modulation
vector, obtained from Ref.
\cite{Vasiliu-Doloc,Kubota}}\label{Fig:4}
\end{figure}

It is still an enigma why $q_{L}$ slowly increases with $x$. The
quasi-linear $x$ dependence of $q_{L}$ reminds us of a simple
$q_{L}=x$ rule, which is experimentally found in the layered
nickelates, which show striped CO correlation
\cite{Tranquada,Yoshizawa}. The stripe-like lattice/charge
correlation above $T_{\rm C}$ in this layered system can be a
generic feature of doped Mott insulators due to the presence of
strong Coulomb repulsion \cite{Tranquada,Yoshizawa}. Then, our
present results suggest an appealing scenario in which \textit{the
charge/lattice correlation synergistically complements the CDW
correlation driven by the nested FS which is formed near and below
$T_{\rm C}$.}

Indeed the FM spin correlation/fluctuation high above $T_{\rm C}$
that coexists with short-range charge/lattice correlation has been
found experimentally \cite{Osborn}. As the short-ranged FM
correlation induces the metallic carriers in the nested FS with
decreasing $T$ toward $T_{\rm C}$, the CDW correlation can become
stabilized with the simultaneous and cooperative amplification of
the charge/lattice correlation of the $q_{L}$ type. This scenario
is consistent with an increased $2\Delta$ as $T$ decreases toward
$T_{\rm C}$ in $x=0.40$.

On further lowering $T$ below $T_{C}$, a significant amount of the
quasi-particles emerge in the nodal direction of the FS while the
band is gapped along the antinodal direction \cite{Mannella}. In
this context, the disappearance of the optical gap below $T_{\rm
C}$ can be understood, since the $k$-dependent dispersion in the
FS is averaged in the optical transition process. Furthermore, the
increased nodal quasiparticles under the FM ordering can also
screen the charge/lattice correlation, resulting in the
disappearance of $2\Delta$.

Note that the $\sigma(\omega)$ of $x=0.30$ and 0.40 in Fig. 1(a)
has a clear pseudo-gap feature even far below $T_{\rm C}$. And, an
occupied band in ARPES spectra yields a pseudo-gap feature, the so
called 'peak-dip-hump' structure in the energy distribution curves
\cite{Mannella}. If we assume that the incoherent absorption peak
in Fig. 1 originates from the transition between the occupied band
to the unoccupied band near FS (See, (ii) of Fig. 3(d)), the
$T$-dependence of the hump peak energies in the ARPES spectra
below $T_{\rm C}$ matches well with that of the incoherent
absorption peak energies in $\sigma(\omega)$, as shown in Fig.
3(c). This observation again corroborates our argument that the
CDW instability should become an essential ingredient of the
incoherent absorption peak, the so called polaron peak, in
$\sigma(\omega)$ as well as the pseudo-gap of the FS. [Note also
that the $T$-dependence of the polaron peak shows the intimate
coupling with the frequency shift of stretching optical phonon
mode \cite{HJ}.]

All of our experimental and theoretical studies coherently suggest
that CDW instability should be incorporated to explain the
enhancement of the $2\Delta$ near $x=0.4$ in LSMO. The enhanced
striped charge/lattice correlation via the CDW instability can
effectively describe the diverse optical, structural, and FS
studies. Our study also suggests that it would be quite
interesting to search for more direct evidence of the CDW
formation, such as nonlinear $I-V$ characteristics. The strong
phonon coupling in the manganite might be another essential aspect
of the intimate link between CDW instability and the various
degrees of freedom, such as spin and orbital.

In conclusion, we have shown experimental evidence that the
optical gap just above the long range spin ordering temperature is
clearly enhanced near $x=0.4$ in
La$_{2-2x}$Sr$_{1+2x}$Mn$_{2}$O$_{7}$. ($0.3 \leq x \leq 0.5$).
This intriguing finding has been explained as the enhancement of
CDW correlation near the specific doping due to the matching of
the $k$-space Fermi-surface nesting vector and the stripe-like
charge/lattice modulation vector. Our experimental findings and
the scenario might be generally applicable to the other layered
transition metal oxides in explaining the anomalous doping
dependent enhancement of the charge/lattice correlation.
\begin{acknowledgments}
We acknowledge B. J. Campbell for discussion. This work was
supported by the CRI program and CSCMR through KOSEF. KHK is
partially supported by the KRF (MOEHRD) (R08-2004-000-10228-0) and
the city of Seoul scientific research initiative. YM is supported
by the Ministry of Education, Science, and Culture, Japan.
\end{acknowledgments}


\end{document}